\documentclass[twocolumn,prl,showpacs]{revtex4}

\begin{document}

\topmargin-1.0cm

\title {Time-dependent Kohn-Sham theory with memory}
\author {H. O. Wijewardane}
\altaffiliation{Present address: Department of Physics and Astronomy, University of Missouri,
Columbia, Missouri 65211}
\author{C. A. Ullrich}
\altaffiliation{Present address: Department of Physics and Astronomy, University of Missouri,
Columbia, Missouri 65211}
\affiliation
{Department of Physics, University of Missouri-Rolla, Rolla, Missouri 65409}
\date{\today}
\begin{abstract}
In time-dependent density-functional theory, exchange and correlation (xc) beyond the adiabatic local 
density approximation can be described in terms of viscoelastic stresses in the electron liquid. 
In the time domain, this leads to a velocity-dependent xc vector potential with a memory containing  
short- and long-range components. The resulting time-dependent Kohn-Sham formalism describes the dynamics of electronic
systems including decoherence and relaxation.
For the example of collective charge-density oscillations in a quantum well,
we illustrate the xc memory effects, clarify the dissipation mechanism, and extract intersubband relaxation rates for weak 
and strong excitations.
\end{abstract}
\maketitle
Time-dependent density-functional theory (TDDFT) \cite{rungegross} has
become a popular tool for describing the dynamics of many-electron systems.
The exact time-dependent exchange-correlation (xc) potential
$v_{\rm xc}[n]({\bf r},t)$ contains information about the previous history of the system, including
its initial state \cite{maitraburke}. However, almost
 all present applications of TDDFT employ the adiabatic approximation for 
$v_{\rm xc}[n]({\bf r},t)$, ignoring
all functional dependence on past time-dependent densities $n({\bf r},t')$, $t'<t$. 
The simplest example is the adiabatic local-density approximation (ALDA):
\begin{equation}\label{alda}
v_{\rm xc}^{\rm ALDA} ({\bf r},t) = \left. \frac{d\epsilon_{\rm xc}(\bar{n})}{d\bar{n}} \right|_{\bar{n} = n({\bf r},t)} ,
\end{equation}
where $\epsilon_{\rm xc}(\bar{n})$ is the xc energy density of a homogeneous electron gas of density $\bar{n}$.
The neglect of retardation in ALDA implies frequency-independent and real xc kernels in linear response \cite{grosskohn}.
This approach has been widely used for calculating molecular excitation energies \cite{casida,FA02}.

There have been several attempts to go beyond the ALDA 
\cite{grosskohn,vignalekohn,VUC,ullrichvignale,dobsonbunner,tokatly}. Vignale and Kohn (VK) \cite{vignalekohn} 
showed that a non-adiabatic {\em local} approximation requires the time-dependent {\em current} ${\bf j}({\bf r},t)$ 
as basic variable, rather than the density $n({\bf r},t)$. This formalism was later cast  in a physically more 
transparent form using the language of hydrodynamics \cite{VUC,ullrichvignale,dobsonbunner,tokatly}, 
where non-adiabatic xc effects appear as viscoelastic stresses in the electron liquid. 

To date, the VK formalism has been applied exclusively in frequency-dependent linear response.
The first application was to calculate linewidths of
intersubband plasmons in semiconductor quantum wells \cite{PRBPRL}.
In the absence of disorder and phonon scattering, ALDA gives infinitely sharp plasmon lines.
The VK formalism includes damping due to electronic many-body effects, in good agreement with experimental linewidths \cite{PRBPRL}.
Van Faassen {\em et al.} \cite{faassen1} calculated static axial polarizabilities 
in molecular chains, which are greatly overestimated with ALDA. For many systems, VK 
led to an excellent agreement with ab initio quantum chemical results. Other recent
studies applied the VK theory to calculate atomic and molecular excitation energies \cite{ullrichburke,faassen2}.

Beyond linear response, a wealth of interesting electron dynamics can be explored using
time-dependent Kohn-Sham (TDKS) theory \cite{rungegross}. This letter presents an analysis and applications of the VK formalism
in the time domain. A striking consequence of the memory and velocity dependence of the
VK xc potential is that it introduces retardation into TDKS theory, which in turn leads to decoherence and energy relaxation. 
This will be illustrated for charge-density oscillations in quantum wells.

Several time-dependent Schr\"odinger equations with dissipation have been proposed in the literature
\cite{kostin,gisin,schuch,scarfone}, using quantized classical frictional forces 
or other phenomenological assumptions. By contrast, the VK xc potential has a microscopic origin,
and satisfies exact conditions such as the harmonic potential theorem \cite{vignalekohn,VUC,ullrichvignale}.

In the presence of external scalar and vector potentials, $v({\bf r},t)$ and
${\bf a}({\bf r},t)$, the TDKS equation is
\begin{eqnarray}\label{tdks}
i\hbar\dot{\varphi}_j({\bf r},t) &=&
\left[ \frac{1}{2m}\left( \hbar \frac{\nabla}{i} + \frac{e}{c} \,{\bf a}({\bf r},t) +
\frac{e}{c} \,{\bf a}_{\rm xc}({\bf r},t)\right)^2
 \right. \nonumber\\
&&{} + v({\bf r},t) + v_{\rm H}({\bf r},t)\bigg] \varphi_j({\bf r},t) \;,
\end{eqnarray}
where $v_{\rm H}$ is the Hartree potential and ${\bf a}_{\rm xc}({\bf r},t)$ is the xc vector potential.
A non-adiabatic, nonlinear xc vector potential has been given in Ref. \cite{VUC} to within second order 
in the spatial derivatives as:
\begin{equation}\label{axc}
\frac{e}{c} \,\dot{a}_{{\rm xc},i}({\bf r},t) = - \nabla_i v_{\rm xc}^{\rm ALDA}({\bf r},t)
+  \sum_j \frac{\nabla_j \sigma_{{\rm xc},ij}({\bf r},t)}{n({\bf r},t)}\:,
\end{equation}
where the viscoelastic stress tensor $\sigma_{\rm xc}$ is defined in terms of the velocity field 
${\bf u}({\bf r},t)={\bf j}({\bf r},t)/n({\bf r},t)$:
\begin{eqnarray}\label{sigmaxc}
\lefteqn{\hspace*{-0.5cm}
\sigma_{{\rm xc},ij}({\bf r},t) = \int_{-\infty}^t dt' \bigg\{
\eta({\bf r},t,t')\bigg[\nabla_j u_i({\bf r},t') + \nabla_i u_j({\bf r},t') } \nonumber\\
&& - \left.\frac{2}{3} \, \nabla \cdot {\bf u}({\bf r},t') \delta_{ij}\right]
+ \zeta({\bf r},t,t') \nabla \cdot {\bf u}({\bf r},t') \delta_{ij} \bigg\}.
\end{eqnarray}
The viscosity coefficients in Eq. (\ref{sigmaxc}) are defined as
\begin{equation}\label{visct}
\eta({\bf r},t,t') = \left. \int\frac{d\omega}{2\pi} \: \tilde{\eta}(\bar{n},\omega) e^{-i\omega(t-t')}
\right|_{\bar{n} = n({\bf r},t)}
\end{equation}
and similar for $\zeta$, where
\begin{eqnarray}\label{visco}
\tilde{\eta}(n,\omega) &=& -\frac{n^2}{i\omega} \, f_{\rm xc}^T \\
\tilde{\zeta}(n,\omega) &=& -\frac{n^2}{i\omega}\left[f_{\rm xc}^L - \frac{4}{3}\,f_{\rm xc}^T - \frac{d^2\epsilon_{\rm xc}}
{dn^2}\right].
\end{eqnarray}
$f_{\rm xc}^L$ and $f_{\rm xc}^T$ are the longitudinal and transverse frequency-dependent xc kernels of
a homogeneous electron gas of density $n$ \cite{grosskohn,qianvignale}.
The apparent ambiguity in Eq. (\ref{visct}) whether the density should be evaluated at $t$ or $t'$
is resolved by noting that the difference involves higher gradient corrections. The same argument 
applies to the difference between the instantaneous position $\bf r$ of a fluid element and its 
``retarded position'' $\bf R$ \cite{dobsonbunner}.

In the following, we consider quantum well systems where all spatial dependence is along the $z$
direction only. One can then transform the xc vector potential, Eq. (\ref{axc}),  into a scalar one:
$v_{\rm xc}(z,t) =  v_{\rm xc}^{\rm ALDA}(z,t) + v_{\rm xc}^{\rm M}(z,t)$ (ALDA+M),
with the memory part given by
\begin{equation} \label{vxcm}
v_{\rm xc}^{\rm M}(z,t) =  - \int_{-\infty}^z \frac{dz'}{n(z',t)}\:
\nabla_{z'} \, \sigma_{{\rm xc},zz}(z',t) \;.
\end{equation}
Assuming that the system has been in the ground state (with zero velocity field) for $t <0$,
the $zz$ component of the xc stress tensor becomes
\begin{equation}
\sigma_{{\rm xc},zz}(z',t) = \int_0^t Y(n(z',t),t-t') \nabla_{z'} u_z'(z',t') dt' \;,
\end{equation}
where the memory kernel $Y$ is given by
\begin{equation}
Y(n,t-t') = \frac{4}{3} \, \eta(n,t-t') + \zeta(n,t-t') \;.
\end{equation}
With the help of the Kramers-Kronig relations for $f_{\rm xc}^L$ we can express the memory kernel as follows:
\begin{equation}
Y(n,t-t') = \frac{2}{3} \,\mu_{\rm xc}
-\frac{n^2}{\pi} \int \frac{d\omega}{\omega} \, \Im f_{\rm xc}^L(\omega) \cos\omega(t-t') ,
\end{equation}
with the xc shear modulus of the electron liquid \cite{qianvignale},
\begin{equation}
\mu_{\rm xc} = \frac{3n^2}{4} \left( \Re f_{\rm xc}^L(0) - \frac{d^2\epsilon_{\rm xc}}{dn^2}\right) .
\end{equation}

\begin{figure}
\unitlength1cm
\begin{picture}(5.0,6.0)
\put(-8.3,-16.7){\makebox(5.0,6.0){
\includegraphics{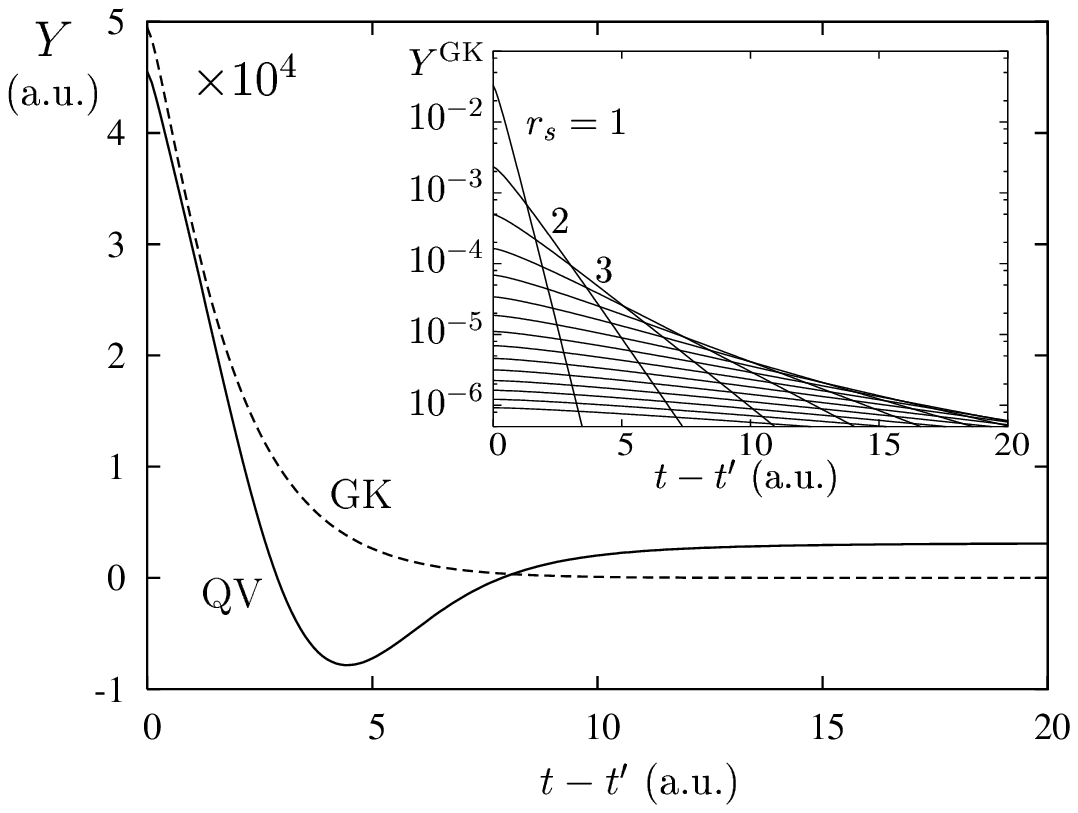}}}
\end{picture}
\caption{\label{figure1} Memory kernel $Y(n,t-t')$ for $r_s=3$, using the QV and GK parametrizations
\cite{grosskohn,qianvignale} for $f_{\rm xc}^L(\omega)$. Inset: $Y^{\rm GK}$ for $r_s$ between 1 and 15, 
indicating exponential memory loss, with a longer-ranged memory for lower densities.}
\end{figure}

Fig. \ref{figure1} shows the memory kernel $Y$ for $r_s=3$, evaluated with the 
Gross-Kohn (GK) \cite{grosskohn} and Qian-Vignale (QV) \cite{qianvignale} parametrizations for $f_{\rm xc}^L(\omega)$.
As shown in the inset, $Y^{\rm GK}$ decreases exponentially over time. The falloff is very rapid for the highest
densities, and the memory becomes more and more long-ranged for lower densities. Comparing the GK and QV parametrizations, 
one finds a similar overall behavior for different values of $r_s$, except that $Y^{\rm QV}$ does not 
decrease monotonically with time but passes through a negative minimum, and then approaches the finite limit $2 \mu_{\rm xc}/3$ 
($\mu_{\rm xc}\to 0$ for large $r_s$  \cite{qianvignale}, while GK assume $\mu_{\rm xc}\equiv 0$ throughout).

To illustrate and analyze the xc memory effects beyond the ALDA, it is convenient to use
a simple analytic model density in Eq. (\ref{vxcm}) to evaluate $v_{\rm xc}^{\rm M}(z,t)$. The function
\begin{equation} \label{modeldensity}
n(z,t) = \frac{2N_s}{L}\:\cos^2 \left(\frac{\pi z}{L}\right)
\left[1 + A \sin\omega t \: \sin\left(\frac{\pi z}{L}\right) \right]
\end{equation}
mimics the non-interacting density of a hard-wall quantum well, 
driven by an AC field of frequency $\omega$, where $|A|\le 1$ to ensure that $n(z,t)\ge 0$.
We take $N_s=1\:a_0^{-2}$ and $L=10\:a_0$ such that $r_s\sim 1$ in the center, and we
simulate a weakly driven case with $\omega=1$ a.u. and $A=0.01$.

\begin{figure}
\unitlength1cm
\begin{picture}(5.0,13)
\put(-9.2,-11.4){\makebox(5.0,13){
\includegraphics{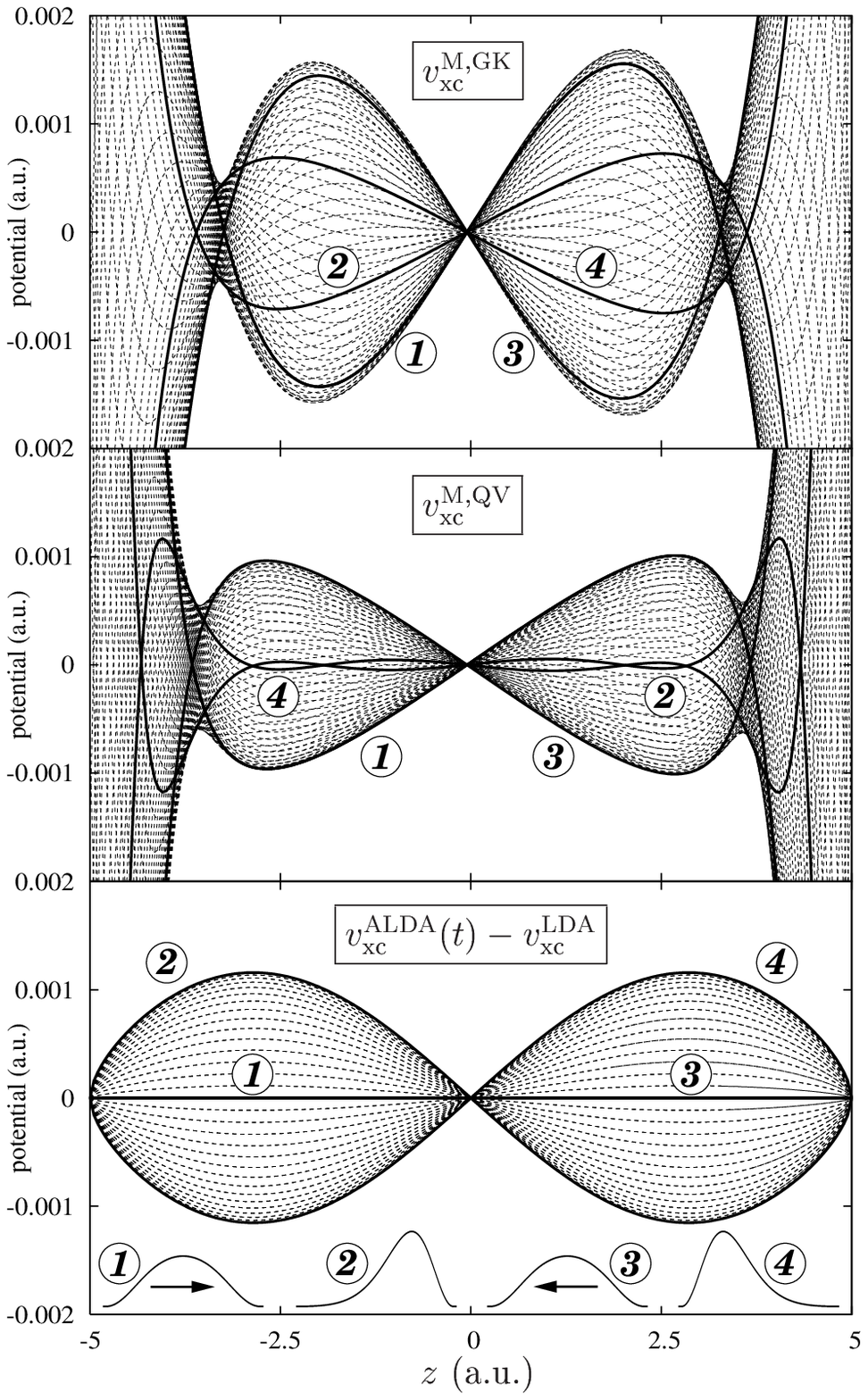}}}
\end{picture}
\caption{\label{figure2} Memory part of the xc potential [Eq. (\ref{vxcm})], evaluated 
for $n(z,t)$ of Eq. (\ref{modeldensity}) in GK and QV parametrization, 
shown in a stroboscopic plot during the 4th cycle after switch-on. 
The heavy lines indicate equidistant snapshots. Compared to the ALDA fluctuations (bottom panel),
$v_{\rm xc}^{\rm M,QV}$ has a phase lag $\pi/2$, and $v_{\rm xc}^{\rm M,GK}$ a little over $\pi/2$. 
The density oscillation in the inset is drawn with enhanced amplitude for clarity. }
\end{figure}

Fig. \ref{figure2} shows a stroboscopic plot of $v_{\rm xc}^{\rm M}(z,t)$ 
during the 4th cycle after switch-on of the time dependent model density.
Four snapshots during the cycle are highlighted: density passes through equilibrium, turns around 
at the right wall, sloshes back, and  hits the left wall.
The reaction of $v_{\rm xc}^{\rm M}$ to these periodic density fluctuations is remarkable:
it opposes the instantaneous current flow by periodically building up an S-like potential barrier 
in the central part of the well, trying to slow down the sloshing motion of the density.
The large-amplitude fluctuations of $v_{\rm xc}^{\rm M}$ close to the edges, 
on the other hand, have little overall impact since they occur in low-density regions.

Both $v_{\rm xc}^{\rm M,GK}$ and $v_{\rm xc}^{\rm M,QV}$ have similar magnitude and shape, and
lag behind the fluctuations of $v_{\rm xc}^{\rm ALDA}$, giving rise to a retardation of the ALDA+M 
xc potential. From the data in Fig. \ref{figure2}, $v_{\rm xc}^{\rm M,QV}$ has a quarter-cycle ($\pi/2$) phase lag, 
which is the most efficient for retardation: the opposing potential barriers are largest at the instants 
of maximal current flow, and become flat whenever the density hits the wall and turns. The good 
synchronization of QV points to a delicate balance between the short- and long-range parts in the memory 
kernel $Y^{\rm QV}$ (see Fig. \ref{figure1}). By comparison, the timing of $v_{\rm xc}^{\rm M,GK}$ is slightly off 
with a phase lag greater than $\pi/2$. As a consequence, 
QV causes stronger damping than GK (see below), even though its potential barriers are lower. 

\begin{figure}
\unitlength1cm
\begin{picture}(5.0,7.5)
\put(-7.7,-12.4){\makebox(5.0,7.5){
\includegraphics{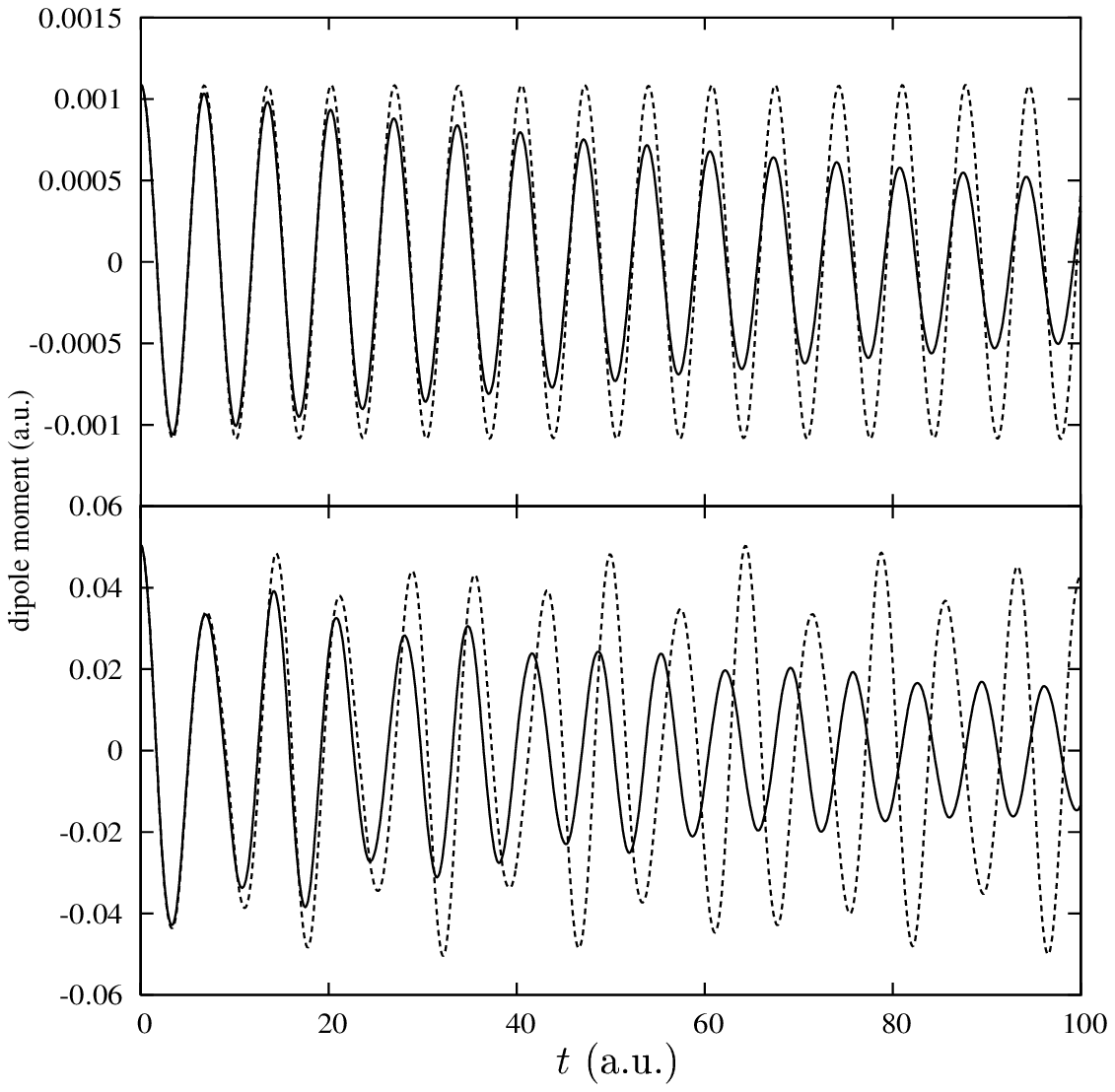}}}
\end{picture}
\caption{\label{figure3} Dipole moment $d(t)$ of a 40-nm GaAs/Al$_{0.3}$Ga$_{0.7}$As quantum well with
electron density $1\times 10^{11} \: \rm cm^{-2}$, initially in a uniform electric field 
${\cal E}_1=0.01 \: \rm mV/nm$ (top) and ${\cal E}_2=0.5\: \rm mV/nm$ (bottom), which is abruptly switched off at $t=0$. 
Dashed lines: ALDA. Full lines: ALDA+M (using QV).}
\end{figure}

We now turn to a more realistic case and solve the TDKS equation for a 40-nm 
modulation-$n$-doped GaAs/Al$_{0.3}$Ga$_{0.7}$As quantum well \cite{PRBPRL,APL},
with effective mass $m^* = 0.067m$, effective charge 
$e^*=e/\sqrt{13}$, well depth 257.6 meV, and electronic sheet density $N_s=1\times 10^{11}\: \rm cm^{-2}$.
The initial condition, obtained from the static Kohn-Sham equation in LDA, 
is the electronic ground state in the presence of a uniform electric field $\cal E$ (``tilted'' 
quantum well). At $t=0$, the electric field is abruptly switched off,
which leaves the quantum well electrons in an excited state and triggers a collective charge-density oscillation.
In the linear regime, this oscillation represents the so-called intersubband (ISB) plasmon, but in the following 
we will also explore the nonlinear, strongly excited regime. Fig. \ref{figure3} shows the dipole moment 
$d(t) = \int\! z\, n(z,t) dz$ versus time ($\rm 1 \: a.u. = 61 \: fs$ in GaAs) for two initial electric fields, 
${\cal E}_1=0.01$ and ${\cal E}_2=0.5$ $\rm mV/nm$, comparing 
ALDA with ALDA+M (using QV; GK gives qualitatively similar results).

Linear-response theory \cite{PRBPRL} yields ISB plasmon frequencies $\omega_{\rm ALDA}=0.9307$,
$\omega_{\rm GK} = 0.9367$, and $\omega_{\rm QV} = 0.9348$ a.u.,
and line(half)widths $\Gamma_{\rm GK} = 0.0061$ and $\Gamma_{\rm QV}=0.0078$ a.u. 
($\rm 1\: a.u. = 10.79 \:meV$ in GaAs). The weak-field case, ${\cal E}_1$, 
is perfectly reproduced by $d(t) = d_0\cos(\omega_{\rm ALDA}t)$ in ALDA 
and $d(t) = d_0\cos(\omega_{\rm QV}t) e^{-\Gamma_{\rm QV}t}$ in ALDA+M. This shows that
$v_{\rm xc}^{\rm M}$  has dissipative and reactive components, giving
an exponential decay of the dipole amplitude with characteristic time $\Gamma_{\rm QV}^{-1}$, and a 
small blueshift of the ISB plasmon frequency.

\begin{figure}
\unitlength1cm
\begin{picture}(5.0,5.4)
\put(-7.5,-15.6){\makebox(5.0,5.4){
\includegraphics{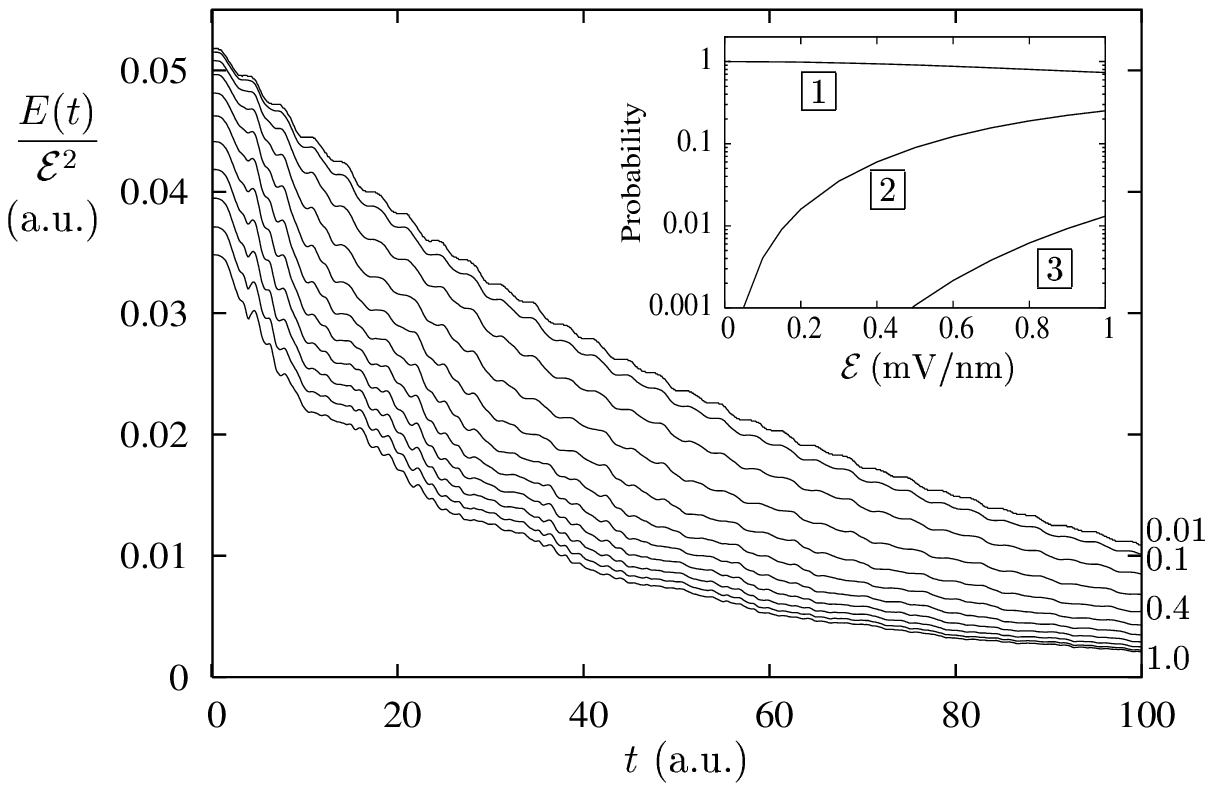}}}
\end{picture}
\caption{\label{figure4} Dissipation of excitation energy $E(t)$,
for the quantum well of Fig. \ref{figure3} and initial fields $\cal E$ between 
0.01 and 1 $\rm mV/nm$ as indicated, calculated with ALDA+M (using QV). Inset: initial occupation
probabilities of the first three subbands.}
\end{figure}

The stronger field, ${\cal E}_2$, causes much richer dynamics. By projecting the initial wavefunction
on field-free Kohn-Sham states, we find 10\% and 0.1\% initial occupation probabilities of the second and third subband levels
(see Fig. \ref{figure4} inset). Consequently, a more complex pattern emerges 
in the dipole oscillations. A spectral analysis shows that $d(t)$ contains higher harmonics of the lowest 
(1-2) ISB plasmon, as well as sidebands resulting from a nonlinear coupling with the dipole-forbidden 1-3 ISB excitation
(the crosstalk between the 1-2 and 1-3 modes is mediated through modulations of the TDKS effective potential).
Due to their larger velocities, the higher-frequency spectral components in $d(t)$ are more rapidly damped 
than the 1-2 ISB plasmon, as seen in Fig. \ref{figure3}.

The decoherence of the dipole oscillations is accompanied by energy dissipation. We define the excitation energy per unit area,
$E(t)$, as the expectation value of the ALDA Hamiltonian with the ALDA+M wave function, minus the ground-state energy. 
Fig. \ref{figure4} shows $E(t)$ scaled by the square of the initial electric field. 
Following the quadratic Stark effect, $E(t)/{\cal E}^2$ is independent of $\cal E$ for small fields $\alt 0.01 \: \rm mV/nm$. 
For larger $\cal E$, higher-order deviations from the quadratic Stark effect emerge.

For small $\cal E$, the excitation energy decreases as $E(t) = E(0)e^{-2\Gamma_{\rm QV}t}$
(small steps are superimposed since the instantaneous dissipation rate depends on the oscillating velocity field
of the ISB plasmon). For larger $\cal E$, one notes deviations from this simple
behavior in the form of a more rapid initial decay and the appearance of
a larger step structure. The origin for these steps is the nonlinear coupling between the 1-2 and
1-3 ISB plasmon modes discussed above, which generates sidebands around $\omega_{\rm QV}^{12}$ whose
frequency spacing $\omega_{\rm QV}^{13}-2\omega_{\rm QV}^{12}$ increases with $\cal E$
(0.27, 0.39, and 0.52 a.u. for 0.1, 0.5, and 1 $\rm mV/nm$). 
As long as these steps are not too pronounced, $E(t)$ is well described by
a biexponential model, with an additional fast channel accounting for the relaxation from higher subbands.
The associated relaxation rate varies between 0.017 and 0.021 a.u. for ${\cal E}$ between 0.1 and $0.3 \: \rm mV/nm$, 
which is more than twice as fast as $\Gamma_{\rm QV}$.

Finally, we comment on the physical mechanism for energy dissipation.
In the linear regime, the VK theory locally assumes a homogeneous electron gas subject to small periodic modulations. 
The frequency-dependent xc kernels then cause the decay of collective modes into multiple particle-hole excitations, 
even if Landau damping is forbidden \cite{vignalekohn,VUC,ullrichvignale}.
In the time domain, one can view the dynamics of an inhomogeneous electron distribution as a superposition 
of local plasmon modes, each subject to decay into multiple particle-hole excitations. 

In conclusion, we have given an explicit demonstration how memory effects 
introduce the element of intrinsic decoherence and energy relaxation into TDKS theory.
This represents an alternative viewpoint to the density-matrix approach \cite{APL}, which, in its simplest form,
describes dissipation through phenomenological decoherence and relaxation times 
(known as $T_1$ and $T_2$ for 2-level systems). A combination of the two approaches suggests itself as a powerful TDDFT tool to 
describe nonlinear electron dynamics in the presence of intrinsic and extrinsic (impurities and disorder) dissipation mechanisms.

\acknowledgments
Acknowledgment is made to the donors of the Petro\-leum Research Fund, administered by the ACS.
C. A. U. is a Cottrell Scholar of the Research Corporation. We thank Giovanni Vignale for fruitful discussions.

\end{document}